\newif\ifanonymous
\newif\ifshort
\newif\iflncs
\lncstrue
\iflncs
\documentclass[11pt]{llncs}
\usepackage{preamble}

\begin{document}
%!TEX root = egalitarianism.tex

\title{Cryptocurrency Egalitarianism:\protect\\A Quantitative Approach}

\iflncs
  \author{
         Dimitris Karakostas \inst{1,3} \and
         Aggelos Kiayias \inst{1,3} \and\\
         Christos Nasikas \inst{2,4} \and
         Dionysis Zindros \inst{2,3}
  }
  \institute{
         University of Edinburgh\and
         University of Athens\and
         IOHK\and
         ``ATHENA'' Research Center
  }
\else
  \titlerunning{Cryptocurrency Egalitarianism}

  \author{Dimitris Karakostas}{University of Edinburgh, IOHK}{dimitris.karakostas@ed.ac.uk}{}{}

  \author{Aggelos Kiayias}{University of Edinburgh, IOHK}{akiayias@inf.ed.ac.uk}{}{}
  \author{Christos Nasikas}{University of Athens, ``Athena'' Research Center}{xnasikas@di.uoa.gr}{}{}
  \author{Dionysis Zindros\footnote{Corresponding author}}{University of Athens, IOHK}{dionyziz@di.uoa.gr}{}{}

  \authorrunning{Karakostas et al.}
\fi
\maketitle

\thispagestyle{plain}
\else
\documentclass[a4paper,UKenglish,cleveref, autoref]{oasics-v2019}
\usepackage{preamble}

\usepackage{longtable}

\bibliographystyle{plainurl}% the mandatory bibstyle

%!TEX root = egalitarianism.tex

\title{Cryptocurrency Egalitarianism:\protect\\A Quantitative Approach}

\iflncs
  \author{
         Dimitris Karakostas \inst{1,3} \and
         Aggelos Kiayias \inst{1,3} \and\\
         Christos Nasikas \inst{2,4} \and
         Dionysis Zindros \inst{2,3}
  }
  \institute{
         University of Edinburgh\and
         University of Athens\and
         IOHK\and
         ``ATHENA'' Research Center
  }
\else
  \titlerunning{Cryptocurrency Egalitarianism}

  \author{Dimitris Karakostas}{University of Edinburgh, IOHK}{dimitris.karakostas@ed.ac.uk}{}{}

  \author{Aggelos Kiayias}{University of Edinburgh, IOHK}{akiayias@inf.ed.ac.uk}{}{}
  \author{Christos Nasikas}{University of Athens, ``Athena'' Research Center}{xnasikas@di.uoa.gr}{}{}
  \author{Dionysis Zindros\footnote{Corresponding author}}{University of Athens, IOHK}{dionyziz@di.uoa.gr}{}{}

  \authorrunning{Karakostas et al.}
\fi
\maketitle

\Copyright{Sara Tucci-Piergiovanni et al.}%TODO mandatory, please use full first names. LIPIcs license is "CC-BY";  http://creativecommons.org/licenses/by/3.0/

% \ccsdesc[100]{Security and privacy}%TODO mandatory: Please choose ACM 2012 classifications from https://dl.acm.org/ccs/ccs_flat.cfm

%TODO mandatory; please add comma-separated list of keywords

\category{}%optional, e.g. invited paper

\relatedversion{}%optional, e.g. full version hosted on arXiv, HAL, or other respository/website
%\relatedversion{A full version of the paper is available at \url{...}.}

\supplement{}%optional, e.g. related research data, source code, ... hosted on a repository like zenodo, figshare, GitHub, ...

\nolinenumbers

\ifanonymous\else
\acknowledgements{Research partially supported by H2020 projects PRIVILEDGE \#780477 and MHMD \#732907.}
\fi

%Editor-only macros:: begin (do not touch as author)%%%%%%%%%%%%%%%%%%%%%%%%%%%%%%%%%%
\EventEditors{Vincent Danos, Maurice Herlihy, Maria Potop-Butucaru, Julien Prat, and Sara Tucci-Piergiovanni}
\EventNoEds{5}
\EventLongTitle{International Conference on Blockchain Economics, Security and Protocols (Tokenomics 2019)}
\EventShortTitle{Tokenomics 2019}
\EventAcronym{Tokenomics}
\EventYear{2019}
\EventDate{May 6--7, 2019}
\EventLocation{Paris, France}
\EventLogo{}
\SeriesVolume{71}
\ArticleNo{5}
%%%%%%%%%%%%%%%%%%%%%%%%%%%%%%%%%%%%%%%%%%%%%%%%%%%%%%

\begin{document}

\maketitle
\fi
%!TEX root = egalitarianism.tex
\begin{abstract}
Since the invention of Bitcoin one decade ago, numerous cryptocurrencies have
sprung into existence. Among these, proof-of-work is the most common
mechanism for achieving consensus, whilst a number of coins have adopted
``ASIC-resistance'' as a desirable property, claiming to be more
``egalitarian,'' where egalitarianism refers to the power of each coin
to participate in the creation of new coins.  While proof-of-work
consensus dominates the space, several new cryptocurrencies
employ alternative consensus, such as proof-of-stake
in which block minting opportunities are
based on monetary ownership.  A
core criticism of proof-of-stake revolves around it being less egalitarian by
making the rich richer, as opposed to proof-of-work in which everyone can
contribute equally according to their computational power. In this paper,
we give the first quantitative definition of a cryptocurrency's
\emph{egalitarianism}. Based on our definition, we measure the egalitarianism of
popular cryptocurrencies that (may or may not) employ ASIC-resistance,
among them Bitcoin, Ethereum, Litecoin, and Monero. Our simulations show,
as expected, that ASIC-resistance increases a cryptocurrency's
egalitarianism.  We also measure the egalitarianism of a stake-based
protocol, Ouroboros, and a hybrid proof-of-stake/proof-of-work cryptocurrency,
Decred. We show that stake-based cryptocurrencies, under correctly selected
parameters, can be perfectly egalitarian, perhaps contradicting folklore belief.
\end{abstract}

%!TEX root = egalitarianism.tex

\section{Introduction}

In 2008, Satoshi Nakamoto proposed Bitcoin~\cite{bitcoin}, the first and most
successful cryptocurrency to date. Bitcoin introduced a cryptographic
consensus protocol in which transactions are organized into
blocks which are put in a globally agreed sequence, the
\emph{blockchain}, despite the presence of adversaries and without the need of
any setup or identity system. Since its inception, a plethora of alternative
cryptocurrencies, or ``altcoins,'' have sprung into existence, each claiming
its own features.

A major thread of blockchain research has focused on %the mandates of block generation,
the mechanics of consensus and 
specifically on the mechanism of identifying the party responsible for producing a
new block at any point. Bitcoin, as well as the majority of altcoins,
employs \emph{proof-of-work}~\cite{C:DwoNao92}, where block generation is
called \emph{mining} and blocks are produced by \emph{miners} who expend
computational power to solve cryptographic puzzles. On the other hand, the most
prominent alternative mechanism is \emph{proof-of-stake}. In proof-of-stake,
block generation is, some times, called \emph{minting} and blocks are produced by
\emph{minters} who ``stake'' their coins, \ie users who own a set of coins and
use them to participate in the consensus protocol. Intuitively, in both cases a
leader is drawn at regular intervals at random from the block generators'
population, with a probability of selection proportional to their computational
power or stake respectively.

Block generators are incentivized to produce blocks by receiving a
\emph{reward} for each block they successfully produce and which is
subsequently adopted in the resulting blockchain.
In many cryptocurrencies, the rewards serve a dual purpose: incentivise the
the miners/minters but also create and distribute the underlying cryptocurrency
to the system's maintainers.
% These rewards follow various schedules that are designed based on the
% macroeconomic desiderada envisioned by the architects of the cryptocurrency. For
% example, the rate of coin production is \emph{halved} every $210\,000$ blocks in
% Bitcoin. Ethereum and Litecoin follow similar schedules. On the contrary,
% Monero has a \emph{smooth emission} schedule in which the rewards are gradually
% reduced at every new block generated. The question of what this schedule should
% be can have significant impact on the variance of stake ownership after an
% execution of a sufficient number of protocol rounds~\cite{equitability}.
Taking this into account, in this paper, we consider the block generators as investors and focus on the
comparison of the \emph{expected} returns of investors with different
purchasing power. The central economic property which arises is that of
cryptocurrency \emph{egalitarianism}. In an ideal world, investing a certain
amount of capital to produce blocks should result in rewards proportional to
that capital; that is, both a \emph{poor} investor and a \emph{rich} investor
should receive returns in proportion to their investment in expectation.
In this point of view,
wealthy investors should not be rewarded
with disproportionate rewards and everybody should have equal opportunity to both participate
and earn rewards. As we will see, this is far from true with most cryptocurrencies
today.

Until now, the term \emph{egalitarianism} has been left undefined, although
several cryptocurrencies claim to be more egalitarian than others \cite{van2013cryptonote} \cite{mcmillan2013}. However,
lacking a quantifiable metric, the question of whether some cryptocurrencies are
more \emph{egalitarian} than others remains ill posed. Our paper aims at
putting forth the first concrete definition of egalitarianism, in a way which is
generic and can be applied to any cryptocurrency.
Our definition provides a metric, which can be practically measured and used
to compare different cryptocurrencies.
Using our model, we measure the egalitarianism of
four indicative proof-of-work--based
cryptocurrencies: Bitcoin, Litecoin~\cite{lee2011litecoin}, Ethereum~\cite{buterin,wood2014ethereum}, and Monero~\cite{van2013cryptonote}. Bitcoin, being the first and most successful cryptocurrency to date, was chosen as
the baseline of comparison. Ethereum is the most promising altcoin and is currently the largest decentralized cryptocurrency by market cap after Bitcoin\footnote{All references to market cap in this paper are made according to \url{https://coinmarketcap.com} [January 2019].}.
Litecoin and Monero, although not next by market cap,
make claims~\cite{van2013cryptonote,mcmillan2013} of increased egalitarianism because of their design.
We assess their claims and find them in agreement with our data, thus presenting for the first time economic comparisons which quantify them precisely.
%  their mining puzzles are
% based on hash functions which are claimed to be memory-hard. Memory-hardness
% has the goal of making it costly to perform large-scale custom hardware attacks
% by requiring large amounts of memory, and hence are claimed to yield more
% egalitarian cryptocurrencies.
On the pure proof-of-stake side, as will soon become clear, egalitarian
behavior is similar across all coins independently of externalities such as hardware characteristics. Therefore,
it suffices to perform a case study of an indicative proof-of-stake protocol.
We study the case of pure proof-of-stake, applied on a protocol consistent with Ouroboros~\cite{C:KRDO17},
as well as a hybrid proof-of-work/proof-of-stake cryptocurrency,
Decred~\cite{decred}.
We find that pure proof-of-stake coins can be perfectly
egalitarian, contrary to their proof-of-work counterparts. However, we
note that variations of proof-of-stake, such as ``delegated proof-of-stake,''
may not be perfectly egalitarian, since the delegates, \ie the leaders of
the stake pools which are formed, typically earn extra profits for managing the
stake pools~\cite{bkks2018}. 
Moreover,  in both cases of proof-of-work and proof-of-stake we consider an {\em open
market} that enables participants to invest in mining or minting without any 
barriers; introducing additional market constraints in acquiring mining 
equipment or stake can similarly disturb the egalitarianism of the underlying system. 

\textbf{Our Contributions and Roadmap.}
This work provides a quantitative evaluation of cryptocurrency egalitarianism.
To the best of our knowledge this is the first work to provide a
treatment of this property and acts as the foundation for comparing
cryptocurrency fairness when it comes to reward distribution.
Specifically, the contributions of our research are summarized as follows:

\begin{enumerate}
  \item We define an exact measure of cryptocurrency
        \emph{egalitarianism}; to do this, we first define the \emph{egalitarian curve} of a
        cryptocurrency from which we extract the measure.
  \item We measure and compare the egalitarian curve and egalitarianism of
        four indicative proof-of-work cryptocurrencies (Bitcoin, Ethereum,
        Litecoin, Monero), one representative proof-of-stake protocol (Ouroboros), and
        a hybrid cryptocurrency (Decred), using current market data.
  \item We show that proof-of-stake, when correctly parameterized, is, perhaps unexpectedly, perfectly
        egalitarian.
\end{enumerate}

The rest of this paper is structured as follows. We begin by reviewing related
work and preliminaries in Sections~\ref{sec:related}
and~\ref{sec:preliminaries}. Next, we put forth our definition for the
egalitarian curve and egalitarianism of a cryptocurrency and motivate its
intuition in Section~\ref{sec:definition}. In Section~\ref{sec:experiments} we
present empirical data for several cryptocurrencies of interest and evaluate
them under our model, in order to deduce whether
previous intuitive claims are indeed correct. Finally, the conclusions of our research
are drawn in Section~\ref{sec:conclusion}.

\section{Related work}\label{sec:related}
The macro and microeconomics of blockchain design have been studied from
several perspectives but remain an active area of research with a number of open questions.
Incentives for block generation
according to the honest protocol have been explored for both proof-of-work and
proof-of-stake.

Proof-of-work protocols such as Bitcoin were formalized in the Bitcoin
Backbone~\cite{EC:GarKiaLeo15,C:GarKiaLeo17} papers and follow-up
works~\cite{pass2017analysis}. The seminal work of Selfish
Mining~\cite{FC:EyaSir14}, see also \cite{FC:SapSomZoh16,kiayias2016blockchain} showed that
honest behavior is not incentive-compatible. Alternative reward sharing mechanisms
in the proof-of-stake  setting
make it feasible to behave better in terms of incentive compatibility for instance
Ouroboros~\cite{C:KRDO17} can be designed from the ground up to be a Nash
equilibrium under certain plausible conditions
and similarly, in the proof-of-work setting~\cite{PODC:PasShi17}.
The question of how to incentivize parties to conduct
pool formation into the desired number of pools, or groups of minters, was
studied in~\cite{bkks2018}.
%
%The above works study the incentives of blockchain systems from the designer's
%point of view so that participants do not deviate from the prescribed protocol.
%A related question is how \emph{fair} the protocol is to participants
%themselves, and in particular to honest participants. The Backbone and Selfish
%Mining works include attacks in which an adversary can strategically harm
%\emph{chain quality}, causing the number of blocks and, in turn, the respective
%rewards, to be disproportionate to their contributed computational power, thereby
%harming fairness against honest participants. Fruitchains
%proposes a protocol which solves this problem. In these works,
%handing out rewards in exact proportion to computational power is considered
%``fair.''

\emph{Egalitarianism} has been studied before in
proof-of-work systems from the perspective of
\emph{memory-hard functions} in~\cite{alwen2017depth,biryukov2016egalitarian},
under the premise that memory hardness provides egalitarianism in the sense
that the it can be used to argue that the
cost of one computational step will be roughly  the same irrespective of the
underlying
computational platform (typically ASIC vs. generic, cf.  \cite{biryukov2016egalitarian}).
The approach we take here instead, asking whether
computational power grows proportionally to capital invested, \ie
whether larger wealth results in more than proportional rewards,
  is more general and it has not
been previously studied to the best of our knowledge.
%Therefore, our work aims at filling this gap by
%studying the effects of economies of scale when applied to cryptocurrency
%generation.

\noindent\textbf{Equitability of cryptocurrencies.}
Fanti \textit{et al.} analyze economic blockchain fairness
in~\cite{equitability}, where they define \emph{equitability}. They study the
evolution of a system after a series of rounds, putting forth the property that
stake ownership remains in proportion \emph{before} and \emph{after} rewards
have been awarded.
By studying the behaviour of the returns' variance under the randomness of
executions, they introduce a geometric reward function and show its optimality
in terms of equitability.  Whereas their equitability metric jointly captures
the normalized variance of rewards for every user conditioned on their initial
resources (e.g., fraction of computational power for PoW), our egalitarianism
metric instead captures the population-wide variation of best-case expected
returns for an initial capital distribution among participants. In other words,
our randomness is over the initial distribution of wealth, whereas theirs is
over the evolution of a single blockchain execution.
In our work, we show that
computational power is not proportional to the invested capital, and hence the
analogy between proof-of-work computational power and proof-of-stake capital
breaks down, and a more detailed study is needed. Additionally, we remark that
proof-of-work miners also reinvest their proceeds in the mining operation,
albeit slowly, as proof-of-stake minters do. For example, empirical data show
that large-scale miners pay for electricity using their
proceeds~\cite{kharif2018}. Hence, both mining and minting follow Pólya
processes as modelled by their paper.
Regardless, \emph{egalitarianism} and \emph{equitability} are orthogonal. A
cryptocurrency can be perfectly egalitarian and poorly equitable and vice versa.
It is possible to obtain a cryptocurrency both egalitarian and equitable by
adopting correctly parameterized proof-of-stake under a geometric reward
function.

\section{Preliminaries}\label{sec:preliminaries}

Before studying the egalitarianism of different cryptocurrency consensus mechanisms, we
provide a description of the leader election process, which is a central part
of each blockchain consensus mechanism. We give an
overview of the details of the two most common decentralized consensus mechanisms,
proof-of-work and proof-of-stake, in order to establish an understanding of the
differences in egalitarianism between the two models.

\noindent\textbf{Proof-of-work.}
The core idea behind proof-of-work cryptocurrencies is solving the
proof-of-work inequality.  Specifically, the mining hardware is provided with
two constants, $\textsc{previd}$ and $\textsc{data}$, \ie the id of the
tip of the adopted blockchain and the data which need to be appended to it.  The mining
device then \emph{brute-force} searches for some string \textsc{nonce}, such
that $\hash(\textsc{previd} || \textsc{data} || \textsc{nonce}) \leq T$ for
some hash function $\hash$ defined by the system. Here, $T$ is a
---relatively--- small number called the \emph{difficulty target}, which is
adjusted in order to ensure a stable block production rate, although typically
remains constant for periods of consecutive blocks called \emph{epochs} --- for
example, in Bitcoin, epochs are $2016$ blocks long~\cite{SP:BMCNKF15}. Because the search for
solutions is exhaustive, the expected number of solutions found by a given
miner is proportional to the number of evaluations of the
hash function $\hash$ she can obtain in a given time frame.

The number of hash evaluations is one of the several critical parameters to
consider when purchasing mining hardware. Other important parameters include
the price of a mining unit, as well as its electricity consumption.  Mining
hardware is divided in various tiers based on performance, namely CPU miners,
GPU miners, FPGA miners, and specialized ASIC miners~\cite{taylor2013bitcoin}.  Although the pricing of
such devices may be similar, the hashing rate and, in turn, the return on
investment, is highly dependent on the hardware's tier.  For example,
the mining hardware ``Whatsminer M10'' produced by the company ``MicroBT''
costs $\$1{,}022.00$ per unit and produces $\$0.104266$ per hour of operation in
net gains, \ie average mined Bitcoins per hour denominated in US dollars with today's
prices (December 2018) minus the electricity costs. On the other hand, the
mining hardware ``8 Nano Pro'' produced by the company ``ASICMiner'' costs
$\$6{,}000.00$ per unit, but produces $\$0.315327$ per hour of operation in net
gains, \ie almost three times the hourly net gains of its cheaper competitor.
Thus, if one can afford to purchase the more expensive hardware, each of their
subsequent dollar invested in electricity returns more mined coins.

It has long been folklore knowledge in the blockchain community that mining
becomes more egalitarian by using a memory-hard proof-of-work function. This
intuition is correct, the core reason being the difficulty to construct
specialized hardware for memory-hard functions. For example, no ASICs
currently exist for Monero mining.  Therefore, the only way to scale mining
operations is by purchasing more general purpose hardware. However, since the mining hardware
in this case varies little, both in terms of cost and performance, scaling
returns become proportional to investments.  To the best of our knowledge, our
work is the first to confirm this correspondence between the
memory-hardness of proof-of-work hash functions and the economics of mining.

\begin{remark}[Block generation at scale]\label{rmk:pow-scale}
We only analyze the scaling of the economics of mining with
respect to hardware. We also do not take into account basic costs such as
shipping and the availability of a basic machine to co-ordinate mining (such as
a personal computer not performing mining itself). A multitude of additional
factors play important roles for mining operations, such as space rental costs,
machine cooling and maintenance costs, as well as bulk electricity purchase.  As
is common in economies of scale, these relative costs are reduced for
large-scale operations, although they are similar for all proof-of-work
cryptocurrencies and thus do not affect relative comparisons between them. We
also remark that we analyze mining costs for small capital investments. If
larger capital, \eg above a few million US dollars, is available, corporations
can develop their own specialized hardware and gain a competitive advantage by
treating it as a trade secret~\cite{taylor2013bitcoin}. Indeed, these details
make our comparison in favour of proof-of-stake \emph{more pronounced}, as
proof-of-stake operations do not incur such types of costs and do not lend
themselves to specialized mining hardware research. We leave the analysis and
calculation of egalitarianism under these parameters for future work.
\end{remark}

\noindent\textbf{Proof-of-stake.}
In proof-of-stake, a minter is selected in proportion to the stake they hold,
which is to say proportionally to the amount of money they own. There exist a
number of flavors of this process. In one case, all coins automatically
participate in the leader election process --- this is the case for Ouroboros~\cite{C:KRDO17} and
Ethereum's Casper~\cite{buterin2017casper}. In a second flavour, the stake has
to \emph{opt-in} to participate in the election by a special process, such as
purchasing a \emph{ticket} or becoming a delegate of the stake of other users.
This is the case for cryptocurrencies such as Decred~\cite{decred} and EOS~\cite{eos}. Among those
participating in the election, a leader is elected at random, in proportion to
their stake.

% While some protocols such as Ouroboros elect \emph{exactly one} leader, there
% are protocols, in both the work and stake setting, such as Bitcoin and Ouroboros
% Praos, which can elect multiple leaders for a particular time slot. However, the
% final blockchain forms a \emph{sequence}, and hence only one block survives among multiple competing blocks.
% As will shortly become clear, our analysis is not sensitive to such nuances.

Proof-of-stake is often criticized for its lack of egalitarianism. The
rationale is that, in proof-of-stake, the more money one stakes, the more money
one generates. Thus, the \emph{rich get richer}, which is precisely the
\emph{opposite} of egalitarianism. Additionally, in proof-of-stake systems, the
money owners could constitute a \emph{closed, rich club}, refusing to share the
assets with any outsiders.  In contrast, this argument claims, proof-of-work is
naturally egalitarian: everyone is paid not according to the money they own,
but according to the computational power they put to work. In this case, since
computational power is a \emph{natural} thing and cannot be exclusively owned,
a closed rich club cannot be formed.
% In the worst case, one can still
% have a minuscule chance of generating a block with computational power by mining
% with pencil and paper~\cite{paper-mining}, or by building one's own computer.
Although this argument seems agreeable at first, the results of our work contradict it.
In fact, correctly parameterized stake-based systems are much more egalitarian than work-based ones.

It is instructive to dispel the above argument intuitively, before we support
our position with data. Firstly, the argument that money can be exclusively
owned, but computational power cannot, is rather misguided. Indeed, this may be true
in the case of a peculiar oligopoly, where a small faction of parties mutually
agrees to never sell to outsiders, despite external demand. However, in an open
market, both money and computational power can be freely purchased and, in
fact, any non-negligible amount of computational power must be necessarily
purchased that way.
In the present work, \textbf{we assume an open market} for both mining hardware
and financial capital which allows participation in the respective systems.
Therefore, given that both money and computational power
are purchasable, we now need to consider the funds one needs to invest either in
technology or in financial capital in order to maximize the returns from a
cryptocurrency's block generation mechanisms. The amount of cryptocurrency
generated by a given investment can be concretely measured and compared, thus
the question can now be analyzed quantitatively and answered concretely.

We should
note that variations of proof-of-stake, such as ``delegated proof-of-stake,''
may not be perfectly egalitarian, since the delegates, \ie the leaders of
the stake pools which are formed, typically earn extra profits for managing the
stake pools~\cite{bkks2018}. In this paper, we only concern
ourselves with non-delegated variants, \ie \emph{pure} proof-of-stake protocols.
We leave the study of the contrast between pool formation mechanism truthfulness
(or Sybil-resilience) and egalitarianism for future work.

%!TEX root = egalitarianism.tex

\section{Defining egalitarianism}\label{sec:definition}

Having established the basics of consensus mechanisms, we now propose the first
definition of an economic measure of \emph{egalitarianism} in cryptocurrencies.
Before we present our definition, let us first state the \emph{desiderata} of such a
definition. First of all, we want to allow concrete measurements to be
performed on cryptocurrencies and data to be extracted in a manner that is
quantitative and not vague. Thus far, the claims for egalitarianism in various
cryptocurrencies have been rather informal, using a rhetoric which fails to include
exact data~\cite{van2013cryptonote,mcmillan2013}. As such, different cryptocurrencies claim egalitarianism over the
others, without demonstrating the claims or provide conclusive arguments.
Secondly, a definition of egalitarianism must measure the protocol maintenance 
returns of a ``rich dollar'' compared to that of a ``poor dollar.''
We thus desire a
measure which, for a particular cryptocurrency, extracts a smaller value
to indicate a \emph{lack of egalitarianism} (\eg a
case where large wealth generates blocks disproportionately faster than
small wealth) and a larger value to indicate \emph{perfect egalitarianism} (where
every invested dollar has exactly equal power in terms of cryptocurrency
generation).

As a means towards establishing our egalitarianism definition, we define the
\emph{egalitarian curve} $f$ of a cryptocurrency. The horizontal axis of this
curve plots the financial capital which is available for investment denominated
in a fiat currency, USD.\footnote{Given that we explore
a small investment duration, it makes little difference whether these are
nominal USD or real USD, as long as they are the same when applying comparisons.}  The
vertical axis plots the Return On Investment (ROI), which measures the
cryptocurrency amount that is freshly generated in the investment period and
remains unspent at the end of the investment period,
given an optimal allocation of the initial capital. We require
the Return On Investment is necessarily \emph{freshly generated}
cryptocurrency; thus, it must be newly mined or minted, and not part of the
initial capital. Of course, purchasing
cryptocurrency which has already been generated is an investment option, but it
is immaterial to our egalitarianism definition, which focuses on measuring the
egalitarianism of freshly generated cryptocurrency.  Finally, the curve is
plotted with a fixed investment duration in mind --- in this paper, we use a
duration of 1 year.  Naturally, curves of different cryptocurrencies can be
compared only if they use the same duration.

\begin{definition}[Egalitarian curve]
    Given a cryptocurrency $c$, an investment period interval $d$, the set of
    all possible investment strategies $\mathcal{B}$, we define the \emph{egalitarian curve}
    $f_{c,d}: \mathbb{R}^+ \longrightarrow \mathbb{R}^+$ of $c$ for
    investment period $d$ as:

    \[
        f_{c,d}(v) = \frac{\underset{B \in \mathcal{B}}{\max}{\mathbb{E}[B(v)]} - v}{v}
    \]
\end{definition}

The value $\underset{B \in \mathcal{B}}{\max}{\mathbb{E}[B(v)]}$ identifies the maximum expectation of
returns across all investment strategies $\mathcal{B}$, \ie the amount of
returns which the \emph{optimal} strategy ensures for a given initial capital $v$.
The expectation is taken with the blockchain execution as a random variable,
since returns vary by execution (the randomness of the execution can affect the
returns of the strategy, as the same strategy can bring larger returns if the
participant is ``lucky'' e.g., it happens to produce many blocks~\cite{equitability}).

We remark first that we do allow strategies to reinvest capital. For instance,
returns earned from mining rewards can be reinvested in electricity costs for
future mining. Furthermore, for unit consistency, we assume the strategy
$B(v)$ returns the freshly generated coins denominated in the same units as the capital $v$ was given in, such that $f$ represents a ROI; thus, we denominate the
generated cryptocurrencies in USD using the market exchange rate.
Second, we assume participants act independently and follow the protocol
according to its specifications. 

Based on the above, it
is now straightforward to define the \emph{ideal egalitarian curve}. In this
case, the ROI is stable regardless of capital invested. Under these ideal
conditions, the amount of freshly generated cryptocurrency is exactly
proportional to the money invested. Thus, the ideal curve is any constant
curve.

% , as depicted in Figure~\ref{fig:ideal}.
%
% \begin{figure}
%     \centering
%     \includegraphics[width=0.4 \columnwidth,keepaspectratio]{figures/ideal.pdf}
%     \caption{The ideal egalitarian curve of an ideal cryptocurrency.}
%     \label{fig:ideal}
% \end{figure}

As an interesting thought experiment, consider the egalitarian curve which is
decreasing. In this case, the poor would receive proportionally more newly
created cryptocurrencies for every dollar they invest, \ie it would be a
redistribution of wealth from the rich to the poor. However, one
can quickly see that, in decentralized cryptocurrencies where the identities of
the participants are unknown, it is impossible to
hope for something better than the constant curve. Indeed, the fact that
decentralized cryptocurrencies allow anonymous generation of new
identities~\cite{douceur2002sybil}
allows a rich investor to split their investment into smaller ones.  Thus, if
the curve were ever to have a negative slope, the sum of the smaller splits of
the rich investment would achieve a higher gain. By the definition of the
curve, which mandates that it depicts the ROI of an \emph{optimal} investment,
this would be a contradiction. The following lemma makes the above intuition
more precise:

\begin{restatable}[Sybil strategies]{lemma}{restateLemSybilStrategies}
\label{lem:sybil}
    Fix a cryptocurrency $c$ and an investment period interval $d$. Given capital $v$,
    for every natural number $i \in \mathbb{N}^\star$, it
    holds that $f_{c,d}(v) \leq f_{c,d}(i \cdot v)$.
\end{restatable}

The proof of this Lemma is available in Appendix~\ref{sec:proofs}.

Using our definition of the egalitarian curve, we now define egalitarianism as
a concrete number. We begin by considering the initial capital $v$ as a random
variable following a certain distribution $\mathcal{D}$. Egalitarianism is
defined as the variance of the expected ROI when the capital is chosen from the
given distribution.

\begin{definition}[Egalitarianism]
  Given a cryptocurrency $c$, an investment period duration $d$ and an initial
  capital distribution $\mathcal{D}$, we define the \emph{egalitarianism} $e$ of $c$
  for investment duration $d$ under initial capital distribution $\mathcal{D}$
  as follows:

  \[
    e_{c,d,\mathcal{D}} = -\textsf{Var}_{v \gets \mathcal{D}}[f_{c,d}(v)]
  \]

  where $f$ is the egalitarian curve of $c$.
\end{definition}

The intuition behind this definition is that, to have egalitarianism, the ROI
must remain the same across different capital investments. As such, any
deviation from the mean is non-egalitarian. Naturally, if the
egalitarianism of a certain cryptocurrency is \emph{higher} than another's, we
say that the former is \emph{more egalitarian} than the latter. Of course, to be
accurate, such comparisons must only be made after fixing the parameters $c$
and $d$ as well as the initial capital distribution $\mathcal{D}$. We will now
fix the distribution $\mathcal{D}$ to be the uniform distribution between a
minimum and a maximum capital. This choice corresponds to the intuition that the
returns are the same for all initial capitals alike. Clearly a cryptocurrency
with an ideal egalitarian curve is perfectly egalitarian, as we now define.

\begin{definition}[Perfect egalitarianism]
  A cryptocurrency $c$ is \emph{perfectly egalitarian} for investment duration
  $d$ and initial capital distribution $\mathcal{D}$ if
  $e_{c,d,\mathcal{D}} = 0$.
\end{definition}

\section{Experimental results}\label{sec:experiments}

Having established our theoretical framework, we now provide experimental
results on the egalitarianism of various cryptocurrencies. Our experiments
utilize the \emph{egalitarian curve} definition of Section~\ref{sec:definition}
in order to concretely confirm --- or disprove --- the egalitarianism claims of
some of the major, both proof-of-work and proof-of-stake, cryptocurrencies.

In conducting our experiments \textbf{we assume a static environment}.
Specifically, we assume that the token prices, as well as the distribution of
funds which are available for purchasing mining hardware are static and follow
the snapshot of the world which we took at the time of writing. Furthermore, we
assume that our mining operation would not substantially affect these
parameters if it were to be applied on this environment. Finally, we assume
that the set of available strategies $\mathcal{B}$ comprises of the honest
strategies, \eg not including selfish mining which could provide better ROI
by diverging from the protocol.

\noindent\textbf{Proof-of-work.}
\noindent
We have experimentally analyzed the egalitarianism of the following
proof-of-work coins: Bitcoin, Litecoin, Ethereum, and Monero. These
cryptocurrencies act as a representative sample among the thousands of existing
cryptocurrencies. Bitcoin is the largest and most successful cryptocurrency by
market cap. Litecoin is the first cryptocurrency aimed at becoming more
egalitarian by replacing Bitcoin's SHA256 work function with scrypt~\cite{percival2016scrypt}, a 
memory-hard function~\cite{DBLP:conf/eurocrypt/AlwenCPRT17}. Ethereum is one of the most promising alternative
cryptocurrencies, the first to support smart contracts, and the second largest
by market cap; its proof-of-work function is different from both Bitcoin and Litecoin.
Finally, Monero is special with claims of strong egalitarianism due to its
memory-hard mining function, Cryptonight~\cite{van2013cryptonote}. Furthermore, its protocol is often
updated to maintain
egalitarianism~\cite{monero-hard-asic}.

As expected, our experiments show that Bitcoin is the least egalitarian of the
four, with Ethereum following next. Monero is more egalitarian than both, with
Litecoin being the most egalitarian among the proof-of-work coins
we have studied. \todo{why?}

For our experimental setting, we worked as follows. First, we collected
empirical data which describe the available mining hardware options in the
market. For each machine choice, we determined the cost of investment. This is
comprised of its initial price (in USD) as well as its energy cost of operation
(in Watts). The cost of operation was translated to USD per hour by considering
the electricity cost of KWh. As a reference, we used the lowest average KWh
cost in the United States, \ie $\$0.08$ per KWh~\cite{energy-cost}. This reference electricity cost is an estimation which
can vary depending on the country of operation.

Second, we use the reported hash rate of each mining hardware machine to
extract an expectation of the freshly mined coins it would generate per hour,
if it were to run continuously. This expectation is taken over the randomness
of all honest blockchain protocol executions. As such, each party is awarded
block rewards in proportion to their computational power. The difference
between revenue per unit of time and cost of operation per unit of time
produces an \emph{income rate}, which is measured in USD per hour.
For our experiments, we use an interval of investment with $|d| = 1$ year. Although
this choice is arbitrary, it corresponds to the usual definition of ROI in
traditional finance.

Our investment strategy is as follows. The
initial available capital is allocated to an upfront technology
investment, in which an integer instance of the Unbounded Knapsack problem
\cite{mathews1896partition} is solved using dynamic programming\footnote{The
source code of our implementation for this calculation is available in our
repository.} to optimize the total cash flow.
Subsequently, as long as the cash flow is positive, the purchased machines
operate for the indicated total duration, reinvesting part of the freshly
minted coins in electricity costs, in order to generate more coins. Eventually,
this strategy produces an income of freshly generated coins, which have not
been spent and are reported as the strategy's income.

To calculate our concrete numbers, we employ the constants shown in
Table~\ref{tbl:work-constants}. We use the expected block generation rates for
each cryptocurrency, as well as the reward per block, token price, and mining
difficulty at the time of writing, all of which we assume remain constant.
The variance of electricity cost, the duration of
investment, as well as small fluctuations in price and difficulty do not qualitatively change the shape of our egalitarian curves (see Appendix~\ref{sec:appendix-qualitative-difference}).
% Although small
% changes in the constants' values will not affect the relative comparison, we
% note that extreme changes may reorder the currencies in terms of
% egalitarianism. For example, consider the case of Bitcoin where all mining
% products, except a relatively cheap one, are unprofitable, a plausible scenario
% in case \eg the price of Bitcoin drops significantly. In this case, Bitcoin
% could become as egalitarian, or even more, as \eg Litecoin. Thus, identifying
% the ``breaking points'' of egalitarianism for each cryptocurrency, compared to
% others, is an interesting problem which needs be considered in future work.

\begin{table}
  \centering
  \resizebox{\textwidth}{!}{%
    \begin{tabular}{|c|c|c|c|c|c|c|c|}
      \hline
      Variable & Description & Unit & BTC & ETH & LTC & XMR & DCR\\
      \hhline{|=|=|=|=|=|=|=|=|}
      $|d|$ & duration of investment & years & \multicolumn{5}{c|}{$1$} \\
      \hline
      $\ec$ & electricity cost & USD / kWh & \multicolumn{5}{c|}{$0.08$} \\
      \hline
      $\bgr(c)$ & block generation rate & blocks / s & 1 / 600 & 1 / 14.7 & 1 / 150 & 1 / 120 & 1 / 298\\
      \hline
      $\thr(c)$ & total hash rate & Thash / s & 34,727,437 & 179.50374 & 174.537 & 0.00033859 & 178,760\\
      \hline
      $\br(c)$ & reward per block & tokens & 12.5 & 3 & 25 & 3.37 & 11.38\\
      \hline
      $\tp(c)$ & token price & token / USD & 4,074.25 & 126.12 & 32.10 & 47.27 & 18.62\\
      \hline
    \end{tabular}
  }
  \caption{A list of the parameters used in our proof-of-work mining simulations. Some parameters are system-agnostic, whereas others depend on the cryptocurrency $c$.}
  \label{tbl:work-constants}
\end{table}

\dionyziz{confirm that changing the total hash rate maintains curve shapes and argue about it in the main text}

Let $\mathcal{M}$ denote the set of
all available mining machines. For each machine $m \in \mathcal{M}$, our empirically
collected data specifies the following parameters:
\begin{inparaenum}[i)]
    \item the energy consumption rate $\ecr(m)$ in Watts,
    \item an initial cost of purchase $\ic(m)$ in USD, and
    \item a hash rate $\hr(m)$ in Terahashes per second.
\end{inparaenum}
Given the above, we can now calculate the expected income rate per hour
$\mathbb{E}[\ir(m)]$ for a given machine $m$ and a cryptocurrency $c$. In the
following equation, the first part identifies the income per hour, \ie the
amount of tokens (denominated in USD) which the machine produces per hour,
whereas the second part of the equation identifies the electricity cost, \ie
the product of the consumed electricity multiplied by the price of a single
KWh:

\[
\mathbb{E}[\ir(m)] = 3600 \cdot \frac{\hr(m)}{\thr(c)} \cdot \br(c) \cdot \bgr(c) \cdot \tp(c) - \ecr(m) \cdot \ec
\]

There are many possible configurations for technology investments. Each
configuration comprises of a number of copies $n \in \mathbb{N}$ of every
machine type $m \in \mathcal{M}$. Therefore, we define each configuration as
$\overline{m} \subseteq \mathcal{M} \times \mathbb{N}$, with
total initial cost of investment for such configuration being
$\ic(\overline{m}) = \sum_{(m, n) \in \overline{m}}{n \cdot \ic(m)}$.

The above figure is given in USD per hour and, since the initial capital should
suffice to buy the machines of the configuration, we require that
$\ic(\overline{m}) \leq v$,
where $v$ is the initial available capital at the beginning of the simulation.

Now, in order to identify the strategy's optimal net income for the
interval $d$, we iterate over all possible machine configurations, for which
the above inequality holds, and choose the one with the maximum returns:

\[
  B_\textsc{OPT}(v)
  =
  \max{
    \{
      \sum_{(m, n) \in \overline{m}}
      {|d|\mathbb{E}[\ir(m)]}:
      \overline{m} \subseteq \mathcal{M} \times \mathbb{N}
      \land
      \ic(\overline{m}) \leq v
    \}
  }
\]

We note that this is only an approximation to the optimal (in our limited model)
solution, which we used in our simulations. We consider this sufficiently close
to optimal to allow for the calculation of egalitarianism. We give an integer
programming formulation of the optimal strategy for capital allocation in
Appendix~\ref{sec:ip}. We remark here that the general problem of mining hardware
allocation (including our simplified approximation) is computationally
hard~\cite{karp1972reducibility}, as both the Knapsack and the Integer
Programming problems are NP-complete.

As the simulation parameters are many and diverse, in order to allow others to
run the experiments with different values, as well as for reasons of
reproducibility and falsifiability, we openly release our mining investment
optimizer as well as our data for public use\ifanonymous\footnote{The link to our mining investment calculator and our mining hardware data,
  which are available under an open source license, has been redacted from this
  version for anonymity purposes.
}\else\footnote{Our mining investment calculator and our mining hardware data are available
  under the MIT license and a Creative Commons 4.0 Attribution License
  respectively at \url{https://github.com/decrypto-org/egalitarianism}.
}\fi.

\begin{figure}
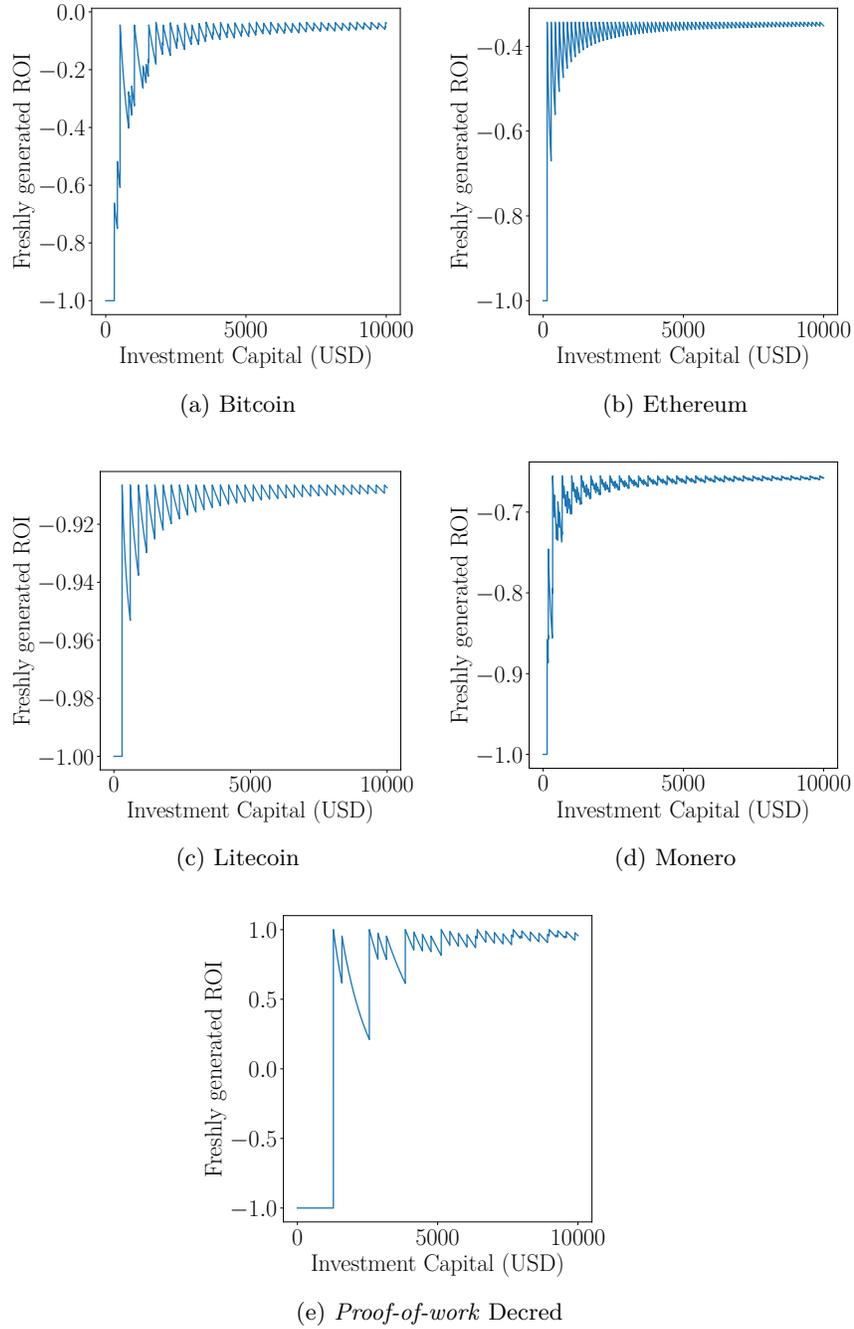

  \placesubfigure{btc_dp_10K_12_months.pdf}{fig:btc_dp_10K_12_months}{Bitcoin}{0.9}{0.5}
  \placesubfigure{eth_dp_10K_12_months.pdf}{fig:eth_dp_10K_12_months}{Ethereum}{0.9}{0.5}
  \placesubfigure{ltc_dp_10K_12_months.pdf}{fig:ltc_dp_10K_12_months}{Litecoin}{0.9}{0.5}
  \placesubfigure{xmr_dp_10K_12_months.pdf}{fig:xmr_dp_10K_12_months}{Monero}{0.9}{0.5}
  \placesubfigure{dcr_dp_10K_12_months.pdf}{fig:dcr_dp_10K_12_months}{\emph{Proof-of-work} Decred}{0.9}{0.5}
  \caption{Egalitarianism curves of the proof-of-work cryptocurrencies analyzed in this work.}
  \label{fig:egalitarian_curves_pow}
\end{figure}

The egalitarianism of Bitcoin, Ethereum, Litecoin and Monero are shown in
Figures~\ref{fig:btc_dp_10K_12_months}, ~\ref{fig:eth_dp_10K_12_months},
~\ref{fig:ltc_dp_10K_12_months}, and~\ref{fig:xmr_dp_10K_12_months}
respectively.
Decred is a hybrid proof-of-work/proof-of-stake cryptocurrency, in
which block generation is a collaboration between miners and minters.
Specifically, each block which is mined via proof-of-work needs to be
``vouched for'' by a certain number of minters, who give it a vote of
confidence. Both the miners and the minters who participate in block
generation are rewarded. An investor can therefore choose to participate in
Decred by either investing in mining hardware and performing proof-of-work, or
by purchasing stake and performing proof-of-stake (or a combination thereof).
We note that the choice of whether to mine or mint Decred is not always clear.
While mining may be more profitable for a certain initial capital, it can also
carry various risks. For instance, if the difficulty increases, the mining
hardware may be rendered inefficient and also hard to sell. Proof-of-work also
carries the operational overhead discussed in Remark~\ref{rmk:pow-scale}. On
the other hand, stake can always be sold, although the price may fluctuate, and
carries negligible operational overhead. As the decision between the two is not
obvious, we analyze both strategies independently. The egalitarianism of
proof-of-work mining for Decred is shown in
Figure~\ref{fig:xmr_dp_10K_12_months}.

It is evident from all figures that the ROI is ``capped'' by a maximum value,
which is observed in specified intervals. Indeed, this value identifies the
ROI of the \emph{best available} machine and is in line with Lemma~\ref{lem:sybil}. In other words, as long as an
investor is able to buy the machine which returns the most profits, then they
achieve the best possible ROI. In case an investor does not have enough capital
to buy the best mining product, they may buy a less profitable machine and
achieve less, though still positive, ROI. This observation explains the small
spikes in ROI which may be seen \eg in Bitcoin's figure for capital in the
range $[0, 2000]$. Also, in case the capital is \emph{more} than the cost of
the machine, then the remaining capital is effectively discarded. Therefore,
although two investors $A, B$ may start with initial capital $v_A < v_B$, if
their returns, in absolute terms, are the same, then the ROI
of $B$ will be smaller as a percentage compared to the ROI of $A$. This
observation explains the decrease in ROI after the spikes. Finally, we observe
that, as the capital increases, the ROI converges to the cap. This is
explained by the fact that the ``discarded'' capital, \ie the capital which
cannot be invested in mining hardware, is a significantly smaller percentage of
the total capital for large investments.

\begin{figure}
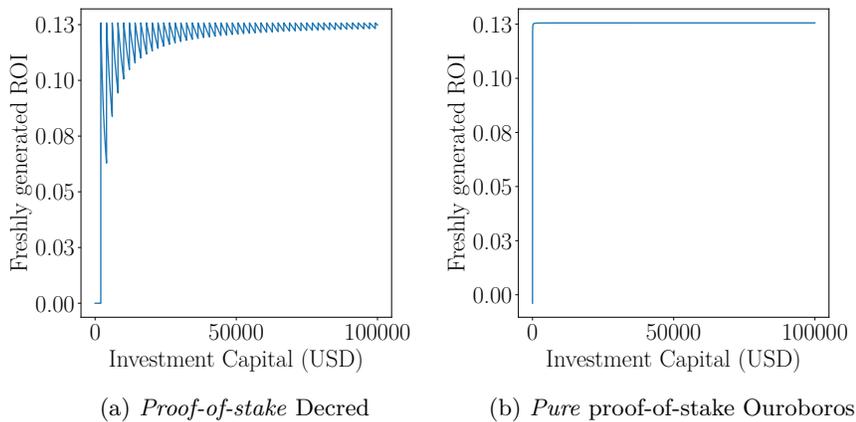

  \placesubfigure{decred-stake.pdf}{fig:decred-stake}{\emph{Proof-of-stake} Decred}{0.9}{0.5}
  \placesubfigure{pure-pos.pdf}{fig:pure-pos-stake}{\emph{Pure} proof-of-stake Ouroboros}{0.9}{0.5}
  \caption{The egalitarianism curves of the proof-of-stake systems analyzed in this work.}
  \label{fig:egalitarian_curves_pos}
\end{figure}

\noindent\textbf{Proof-of-stake.}
\noindent
We now analyze the proof-of-stake egalitarianism in two settings. First, we
consider pure proof-of-stake, which can be applied on top of a protocol like
Ouroboros. In this case, \emph{pure} is in opposition to \emph{delegated}
proof-of-stake, a setting where the stakeholders are required to delegate their
stake to other parties, namely ``stake pools'' and is deployed in
cryptocurrencies such as EOS, Bitshares, and others. Second, we consider the
case of minting Decred via its proof-of-stake mechanism.

The egalitarian curve for \emph{staking} Decred is illustrated in
Figure~\ref{fig:decred-stake}.
% The curve is almost perfectly egalitarian.
As mentioned above, Decred is an opt-in staking cryptocurrency, where staking
occurs by purchasing so-called \emph{tickets}. Since the price of a ticket is
quantized, egalitarianism is harmed for capitals which are not multiples of
ticket prices. However, one can see that the envelope of maxima of this curve
is perfectly egalitarian. The spikes that cause the discontinuity of the curve
are due to the large ticket price (currently $\$1756$), which in Decred is
determined by the market and is high due to the limited supply of tickets
available per ticket pool, a parameter inherent in their protocol. Perfect
egalitarianism could in principle be achieved by making the ticket price
approach $0$.

In the case of Ouroboros, every coin has the same probability of
being chosen for extending the chain~\cite{C:KRDO17}. When a coin is eligible for block
generation, its owner can create a block by providing a proof of ownership of
the chosen coin. Consider the case of a cryptocurrency with $N$ coins in
circulation. When a block needs to be created, a coin is chosen at random from
the set of $N$ coins. Therefore, each coin may be chosen with $1 \over N$
probability. Then the address which owns the chosen coin, in other words the
stakeholder which controls this coin, is eligible to generate a block and
receive the block rewards associated with it.\footnote{In~\cite{C:KRDO17} 
the payout does not explicitly include 
 freshly minted coins and is comprised of transaction fees. We consider an identical
 reward schedule as~\cite{C:KRDO17} but comprised only of freshly minted coins.}  In our experiments, we assume that every block is associated with a
constant reward, which pertains to newly minted coins. Furthermore, since
computational power does not affect the rate of block production, it is
reasonable to assume that both the electricity and the hardware equipment's
price is constant for all users, regardless of stake accumulation, so all users
can participate using --- relatively --- cheap resources (cf. Remark~\ref{rmk:pow-scale}).

Figure~\ref{fig:pure-pos-stake} depicts the simulation of a pure proof-of-stake
system. In this case, the users pay a set transaction fee\footnote{As of January $2019$, according to
\url{https://cardanoexplorer.com/}, in the Cardano implementation of Ouroboros
these fees are in the order of $\$0.01$.}  for the
purchase of the initial stake. The rest of their capital is allocated as stake.
The figure suggests that this system is closer to perfect egalitarianism
compared to the rest of our case studies.

\noindent\textbf{Summary.}
\noindent
Our findings are summarized in Table~\ref{tbl:egalitarianism}. We find that
Bitcoin is the least egalitarian, followed in turn by Ethereum, Monero, and
Litecoin\footnote{Litecoin may appear to have better egalitarianism
compared to Monero due to limited availability of mining machines.
More data are needed to economically compare scrypt and
CryptoNight mining.}. The latter two are the
most egalitarian due to their use of CryptoNight and scrypt respectively. Mining
with Decred provides the worst egalitarianism of all tested coins. \todo{why?}
However, the most egalitarian coins involve staking. Decred staking, due to its
quantized ticket pricing, is only approximately egalitarian and comparable to
the performance of mining Litecoin. Pure proof-of-stake, which allows continuous
staking, is \emph{almost perfectly} egalitarian, its small divergence from
perfect egalitarianism stemming from the small capital which is required to pay
the transation fees to participate in the staking process.

\begin{table}
  \centering
  \begin{tabular}{|c|c|c|}
    \hline
    Name & Consensus mechanism & Egalitarianism \\
    \hhline{|=|=|=|}
    Bitcoin &  Proof-of-work &  -0.034490298 \\
    \hline
    Ethereum & Proof-of-work &  -0.006926114 \\
    \hline
    Litecoin & Proof-of-work &  -0.000271822 \\
    \hline
    Monero &   Proof-of-work &  -0.002206135 \\
    \hline
      Decred & \begin{tabular}{c} Proof-of-work \\ Proof-of-stake \end{tabular} & \begin{tabular}{c} -0.412524642 \\ -0.000348280 \end{tabular}  \\
    \hline
    Ouroboros & Proof-of-stake & -0.000000295 \\
    \hline
  \end{tabular}
  \caption{A comparison of the egalitarianism values of the cryptocurrencies explored in this study.}
  \label{tbl:egalitarianism}
\end{table}

\section{Conclusion}\label{sec:conclusion}

In this work we explore the notion of \emph{egalitarianism} of
cryptocurrencies. Although this notion has long been discussed, we are the first
to give a definition, which allows us to concretely argue about the
egalitarianism of various existing systems.

The results of our experimental simulations are very optimistic in
terms of usability of our metric, as they provide concrete figures
which measure the egalitarianism of several popular cryptocurrencies.
The most unexpected result arises from the comparison between the proof-of-work
and proof-of-stake mechanisms. Although blockchain folklore argued in favour of
proof-of-work systems in terms of egalitarianism,
our results show that, in fact, it is proof-of-stake systems which are more
egalitarian in our model.

Our work provides the first step towards establishing a concrete framework of
egalitarianism evaluation in the cryptocurrency ecosystem. Future work will
focus in evaluating more existing cryptocurrencies and
investigating variations of consensus mechanisms such as delegated
proof-of-stake. Additionally, we leave for future work the treatment of more
complex economical models
of the mining game such as dynamic systems and adversarial strategies, as well
as economies of scale in the multitude of parameters we have ignored, such as
electricity bulk pricing. We conjecture the consideration of such parameters
will exacerbate the gap between proof-of-work and proof-of-stake which we have
illustrated in this work.

Finally, we remark that neither proof-of-work nor proof-of-stake blockchains
are politically egalitarian systems, in which the ideal of
\emph{one human one vote} is attained. Instead, at best,
\emph{one coin one vote} is attained in the case of well-parameterized
proof-of-stake systems. Thus, as illustrated in this paper, blockchain systems are, for
the time being, plutocratic. Whether decentralized decision making in which
each human is allocated one vote is possible remains an open question.
In such a system, the egalitarian curve would be strictly
decreasing; however, our results, especially Lemma~\ref{lem:sybil}, hint towards our
conjecture that such systems are impossible in a Sybil-resilient setting.

\iflncs
\bibliographystyle{plain}
\fi
\bibliography{pubs,abbrev3,crypto}

\newpage
\appendix

\section{Proofs}\label{sec:proofs}

\restateLemSybilStrategies*
\begin{proof}
    We prove the statement via contradiction. Assume that for capital $v$
    exists a natural number $i \in \mathbb{N}^\star$ such that
    $f_{c,d}(v) > f_{c,d}(i \cdot v)$. Also assume that for
    capital $v$ the optimal strategy is $B'$, so: $\underset{B \in
    \mathbb{B}}{\max}{\mathbb{E}[B(v)]} = \mathbb{E}[B'(v)]$. Then, for capital
    $i \cdot v$ exists a strategy $B''$, such that the capital is split into $i$
    equally-sized parts and the strategy $B'$ is applied on each part. Given
    that the executions of the substrategies on these parts are independent,
    then the expected returns for the strategy $B''$ are:
    \begin{align}\label{eq:break-strategy}
        \mathbb{E}[B''(i \cdot v)] = i \cdot \mathbb{E}[B'(v)]  = i \cdot \underset{B \in \mathbb{B}}{\max}{\mathbb{E}[B(v)]}
    \end{align}
    It also holds that $B''$ is at best the optimal strategy, so:
    \begin{align}\label{eq:multi-strategy}
        \underset{B \in \mathbb{B}}{\max}{\mathbb{E}[B(i \cdot v)]} \geq \mathbb{E}[B''(i \cdot v)] \xRightarrow{\text{(\ref{eq:break-strategy})}}
        \underset{B \in \mathbb{B}}{\max}{\mathbb{E}[B(i \cdot v)]} \geq i \cdot \underset{B \in \mathbb{B}}{\max}{\mathbb{E}[B(v)]}
    \end{align}
    However, it should hold that:
    \begin{alignat}{2}
        f_{c,d}(v) &> f_{c,d}(i \cdot v) \Rightarrow \notag\\
        \frac{\underset{B \in \mathbb{B}}{\max}{\mathbb{E}[B(v)]} - v}{v} &> \frac{\underset{B \in \mathbb{B}}{\max}{\mathbb{E}[B(i \cdot v)]} - i \cdot v}{i \cdot v} \xRightarrow{\text{(\ref{eq:multi-strategy})}} \notag\\
        \frac{\underset{B \in \mathbb{B}}{\max}{\mathbb{E}[B(v)]} - v}{v} &> \frac{i \cdot \underset{B \in \mathbb{B}}{\max}{\mathbb{E}[B(v)]} - i \cdot v}{i \cdot v} \Rightarrow \notag\\
        \frac{\underset{B \in \mathbb{B}}{\max}{\mathbb{E}[B(v)]} - v}{v} &> \frac{\underset{B \in \mathbb{B}}{\max}{\mathbb{E}[B(v)]} - v}{v}
    \end{alignat}
    which is impossible.
\end{proof}

\section{Integer programming formulation}\label{sec:ip}

In our experiments, we used a Dynamic Programming solution to solve the Knapsack
problem in order to allocate mining machines upfront. An optimal solution could
use the proceeds of mining not only to reinvest in electricity, but also to
purchase new machines. This is captured by the Integer Programming
formulation in Figure~\ref{fig:IP}, which gives the optimal investment strategy
in the full model.

This maximization problem tries to optimize the \emph{freshly} generated
proceeds. The variables to solve for, $x_{m,t} \in \mathbb{N}$, describe the
number of machines of type $m$ that the investor holds at time $t$. We assume
machines cannot be sold back to the market, hence $x_{m,t-0} \leq x_{m,t}$. The
investment starts with initial capital $v$ and no machines, hence $x_{m,0} = 0$.
The program can then decide to purchase machines as time goes by.
For any costs, it first uses up the initial capital $v$ to pay for
them (this initial capital is useless to keep, as it does not contribute to
freshly generated proceeds, which are our utility here), and subsequently uses
the proceeds to pay for any remaining costs. Capital which is not expended to
pay for costs is discarded by the $\max$ operator in the maximization clause.
The condition the integer program is subject to requires that the investment has
non-negative capital at every point in time, and hence does not run out of
money. In this formulation, it is assumed that $d$ is a set of consecutive
integers representing indexed \emph{hours} of execution (a more fine-grained
solution can be obtained by increasing this temporal resolution as needed).

\begin{figure}
  \noindent\fbox{%
      \parbox{\textwidth-2ex}{%
        \begin{align*}
        &\textbf{Maximize }\\
           &\qquad \max(
             0,
            v - \sum_{t \in d \setminus d[0]}
                   \sum_{m \in \overline{m}}
                     (x_{m,t} - x_{m,t-1}) \ic(m_i)
              - \sum_{t \in d}
                   \sum_{m \in \overline{m}}
                      x_{m,t} \cdot \ecr(m) \cdot \ec
           )
           \\
           &\qquad +
           \sum_{t \in d}
             \sum_{m \in \overline{m}}
               x_{m,t} \cdot
               3600 \cdot \frac{\hr(m)}{\thr(c)} \cdot \br(c) \cdot \bgr(c) \cdot \tp(c)
           \\
        &\textbf{subject to  }\\
          &\qquad \sum_{\substack{t' \leq t\\t' \neq d[0]}}
            \sum_{m \in \overline{m}}
               (x_{m,t'} - x_{m,t'-1}) (
               -\ic(m_i)
               +(t - t' + 1)|\mathbb{E}[\ir(m)]
                )
          \leq v
          \text{ for } t \in d\\
          &\qquad x_{m,t-1} \leq x_{m,t} \text{ for } m \in \overline{m} \text{ and } t \in d \setminus d[0]\\
          &\qquad x_{m,d[0]} = 0 \text{ for } m \in \overline{m}\\
        &\textbf{and }
          x_{m,t} \in \mathbb{N} \text{ for } m \in \overline{m} \text{ and } t \in d
        \end{align*}
      }%
  }
  \caption{An Integer Programming formulation of the optimal investment strategy in our model.}
  \label{fig:IP}
\end{figure}

\section{Parameters affecting egalitarianism}\label{sec:appendix-qualitative-difference}

\begin{figure}
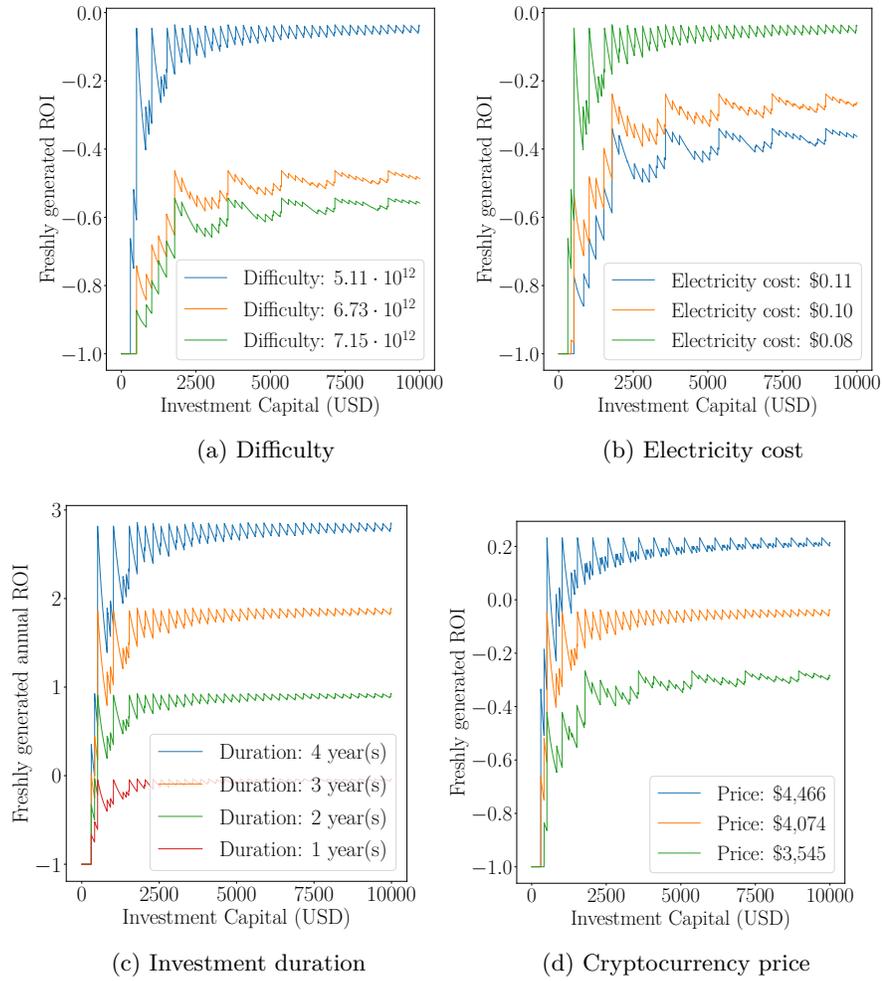

  \placesubfigure{btc_dp_10K_diff_difficulty.pdf}{fig:different_difficulty}{Difficulty}{0.9}{0.5}
  \placesubfigure{btc_dp_10K_diff_kwh.pdf}{fig:different_electricity_cost}{Electricity cost}{0.9}{0.5}
  \placesubfigure{btc_dp_10K_diff_time.pdf}{fig:different_duration}{Investment duration}{0.9}{0.5}
  \placesubfigure{btc_dp_10K_diff_rate.pdf}{fig:different_price}{Cryptocurrency price}{0.9}{0.5}
  \caption{Bitcoin egalitarian curves under varying parameters.}
  \label{fig:different-settings}
\end{figure}

Throughout this paper, we have assumed certain parameters (cryptocurrency
prices, electricity prices, duration of investment and mining difficulty) remain
constant throughout the investment period. Furthermore, we have taken into account
current market values to the best of our knowledge. We note that, while the
actual egalitarianism numbers may change depending on these parameters, the
general shape of egalitarian curves and the qualitative comparison between
different cryptocurrencies remains the same. To illustrate this point, we have
measured the egalitarian curve of Bitcoin for varying parameter values. Our
results are demonstrated in Figure~\ref{fig:different-settings}.

\section{Machines}\label{sec:appendix-data}

Data for mining machines was obtained from a multitude of resources on the
Internet\footnote{An exhaustive list of our resources includes the online
stores \url{https://whattomine.com/}, \url{https://cryptominer.deals/},
\url{https://www.asicminervalue.com/},
 \url{https://www.reddit.com/r/MoneroMining/comments/9omjfb/rtx_2080_ti_mining_monero_at_1228hs_and_more/},
\url{https://www.newegg.com/}, \url{https://www.amazon.com/},
\url{https://shop.bitmain.com.cn}, \url{https://www.cryptouniverse.at},
\url{https://canaan.io}, \url{http://miner.ebang.com.cn},
\url{https://swminershop.com}, \url{https://asicminer.co},
\url{https://estrahash.com}, \url{http://www.innosilicon.com},
\url{https://pangolinminer.com}, \url{https://www.bitfily.io},
\url{https://hashdeploy.net/}, \url{https://www.pantech.company},
\url{https://www.cryptominerbros.com}, \url{https://pandaminer.com},
\url{https://minersdeals.com}, \url{https://sharkmining.com},
\url{https://shop.miningstore.com}, \url{https://mineshop.eu},
\url{https://www.bitmart.co.za}, \url{https://shop.futurebit.io},
\url{https://www.aliexpress.com}, \url{https://bitech-mining.com},
\url{https://asicminermarket.com}, \url{https://www.baikalminer.com},
\url{https://prominerz.com}}. Data for graphics processing units (GPU) and
central processing units (CPU) was obtained by calculating the median of
multiple user benchmarks when available\footnote{\url{https://www.xmrstak.com/tag/monero/}, \url{https://gpustats.com/},
\url{https://www.ethmonitoring.com/benchmark},
\url{https://monerobenchmarks.info/}}. The price of each machine used in our experiments is the
reported retails price of machine at date of access. When a new machine is not available
for sale, the price of a used or refurbished machine is used. For reproducibility
purposes, our complete data set is openly available in our repository. For
reference, we list a summary of those machines which provide a
\emph{positive net gain per hour} after purchase (and can thus be profitable
under our assumed parameter values) in Table~\ref{tbl:machines}.

\begin{longtable}{|p{.60\textwidth}|r|r|r|}
  \hline
  \multicolumn{4}{|c|}{\textbf{Bitcoin}} \\
  \hline
  Name & Hashes / s & Watt & Price (USD) \\
  \hhline{|=|=|=|=|}
  8 Nano Pro & $76 \cdot 10^{12}$ & 4,000 & 6,000 \\
  Whatsminer M10S & $55 \cdot 10^{12}$ & 3,500 & 2,558 \\
  Ebit E11++ & $44 \cdot 10^{12}$ & 1,980 & 2,024 \\
  8 Nano & $44 \cdot 10^{12}$ & 2,100 & 1,790 \\
  T3 43T & $43 \cdot 10^{12}$ & 2,100 & 2,279 \\
  Ebit E11+ 37 & $37 \cdot 10^{12}$ & 2,035 & 1,517 \\
  WX6 & $34 \cdot 10^{12}$ & 3,200 & 1,275 \\
  Whatsminer M10 & $33 \cdot 10^{12}$ & 2,145 & 1,022 \\
  T2T-32T & $32 \cdot 10^{12}$ & 2,200 & 1,568 \\
  Ebit E11 & $30 \cdot 10^{12}$ & 1,950 & 1,110 \\
  Antminer S15 (28T) & $28 \cdot 10^{12}$ & 1,596 & 1,249 \\
  Antminer S15 (27T) & $27 \cdot 10^{12}$ & 1,539 & 1,363 \\
  T2T-25T & $25 \cdot 10^{12}$ & 2,050 & 1,150 \\
  Snow Panther B1+ & $24.5 \cdot 10^{12}$ & 2,100 & 580 \\
  T2T-24T & $24 \cdot 10^{12}$ & 1,980 & 1,350 \\
  S11i & $24 \cdot 10^{12}$ & 2,300 & 937 \\
  Antminer T15 & $23 \cdot 10^{12}$ & 1,541 & 840 \\
  Antminer S11 & $20.5 \cdot 10^{12}$ & 1,435 & 512 \\
  AvalonMiner 921 & $20 \cdot 10^{12}$ & 1,800 & 415 \\
  Antminer S9-Hydro & $18 \cdot 10^{12}$ & 1,728 & 713 \\
  Ebit E10 & $18 \cdot 10^{12}$ & 1,650 & 2,999 \\
  T2 Terminator & $17.2 \cdot 10^{12}$ & 1,570 & 1,118 \\
  DragonMint T1 & $16 \cdot 10^{12}$ & 1,480 & 1,600 \\
  AvalonMiner 851 & $15 \cdot 10^{12}$ & 1,450 & 380 \\
  Antminer S9i & $14.5 \cdot 10^{12}$ & 1,365 & 440 \\
  Antminer S9j & $14.5 \cdot 10^{12}$ & 1,365 & 307 \\
  AvalonMiner 841 & $13.6 \cdot 10^{12}$ & 1,290 & 354.44 \\
  SX6i & $11 \cdot 10^{12}$ & 900 & 419 \\
  \hline
  \multicolumn{4}{|c|}{\textbf{Ethereum}} \\
  \hline
  Name & Hashes / s & Watt & Price (USD) \\
  \hhline{|=|=|=|=|}
  A10 EthMaster & $485 \cdot 10^{6}$ & 850 & 5,399 \\
  A10 EthMaster & $432 \cdot 10^{6}$ & 740 & 4,799 \\
  \begin{tabular}{@{}l@{}}Shark Extreme 2\\\hspace{1cm}\small(8$ \times $NVIDIA GTX 1080 Ti)\end{tabular}& $420 \cdot 10^{6}$ & 1,500 & 9,779 \\
  Maximus+ \small(8$ \times $1080TI) & $370 \cdot 10^{6}$ & 2,200 & 7,520 \\
  A10 EthMaster & $365 \cdot 10^{6}$ & 650 & 4,099 \\
  \begin{tabular}{@{}l@{}}Ethereum Mining Rig\\\hspace{1cm}\small(12x AMD RX 570 GPU)\end{tabular} & $360 \cdot 10^{6}$ & 1,600 & 4,345 \\
  ULTRON \small(8$ \times $P104) & $320 \cdot 10^{6}$ & 1,700 & 5,338 \\
  \begin{tabular}{@{}l@{}}Ethereum Mining Rig\\\hspace{1cm}\small(8$\times$ NVIDIA 1080 8GB GPU)\end{tabular} & $310 \cdot 10^{6}$ & 1,100 & 6,267 \\
  \begin{tabular}{@{}l@{}}Shark Extreme 2\\\hspace{1cm}\small(6$ \times $NVIDIA GTX 1080 Ti)\end{tabular} & $300 \cdot 10^{6}$ & 1,200 & 7,880 \\
  Shark Extreme 2 \small(8$ \times $AMD Vega 56) & $290 \cdot 10^{6}$ & 1,700 & 6,879 \\
  \begin{tabular}{@{}l@{}}Shark Extreme 2\\\hspace{1cm}\small(8$ \times $NVIDIA GTX 1070 Ti 8 GB)\end{tabular} & $245 \cdot 10^{6}$ & 1,400 & 6,679 \\
  Shark Extreme 2 \small(8$ \times $AMD RX 580) & $240 \cdot 10^{6}$ & 1,100 & 4,590 \\
  \begin{tabular}{@{}l@{}}Ethereum Mining Rig\\\hspace{1cm}\small(8$\times$AMD MSI RX 580 GPU)\end{tabular} & $240 \cdot 10^{6}$ & 1,000 & 3,453 \\
  IMPERIUM+ \small(8$ \times $RX 570/580) & $230 \cdot 10^{6}$ & 1,300 & 3,577 \\
  Antminer G2 & $220 \cdot 10^{6}$ & 1,200 & 3,799 \\
  Shark Extreme 2 \small(6$ \times $AMD Vega 56) & $220 \cdot 10^{6}$ & 1,275 & 5,680 \\
  \begin{tabular}{@{}l@{}}Ethereum Mining Rig\\\hspace{1cm}\small(8$\times$AMD MSI RX 570 GPU)\end{tabular} & $220 \cdot 10^{6}$ & 950 & 3,2253 \\
  \begin{tabular}{@{}l@{}}Shark Extreme 2\\\hspace{1cm}\small(4$ \times $NVIDIA GTX 1080 Ti)\end{tabular} & $210 \cdot 10^{6}$ & 800 & 4,979 \\
  Antminer E3 & $190 \cdot 10^{6}$ & 760 & 654 \\
  \begin{tabular}{@{}l@{}}Shark Extreme 2\\\hspace{1cm}\small(6$ \times $NVIDIA GTX 1070 Ti 8 GB)\end{tabular} & $185 \cdot 10^{6}$ & 1,050 & 5,480 \\
  Shark Extreme 2 \small(6$ \times $AMD RX 580) & $180 \cdot 10^{6}$ & 825 & 3,890 \\
  \begin{tabular}{@{}l@{}}Ethereum Mining Rig\\\hspace{1cm}\small(6$\times$AMD RX580 8gb GPU)\end{tabular} & $180 \cdot 10^{6}$ & 900 & 2,342 \\
  \begin{tabular}{@{}l@{}}Ethereum Mining Rig\\\hspace{1cm}\small(6$\times$AMD MSI RX 580 GPU)\end{tabular} & $175 \cdot 10^{6}$ & 860 & 1,967 \\
  \begin{tabular}{@{}l@{}}Ethereum Mining Rig\\\hspace{1cm}\small(6$\times$AMD MSI RX 580 GPU)\end{tabular} & $170 \cdot 10^{6}$ & 750 & 2,156 \\
  Thorium 6580 GPU & $160.2 \cdot 10^{6}$ & 700 & 4,297 \\
  Thorium 6570 GPU & $144 \cdot 10^{6}$ & 750 & 3,974 \\
  \begin{tabular}{@{}l@{}}Shark Extreme 2\\\hspace{1cm}\small(4$ \times $NVIDIA GTX 1070 Ti 8 GB)\end{tabular}& $122 \cdot 10^{6}$ & 600 & 3,580 \\
  Zodiac 6-1060 GPU & $120.78 \cdot 10^{6}$ & 750 & 3,222 \\
  Shark Extreme 2 \small(4$ \times $AMD RX 580) & $120 \cdot 10^{6}$ & 550 & 2,590 \\
  \begin{tabular}{@{}l@{}}Ethereum Mining Rig\\\hspace{1cm}\small(6$\times$AMD MSI RX 560)\end{tabular} & $80 \cdot 10^{6}$ & 370 & 1,823 \\
  GeForce RTX 2080Ti & $55 \cdot 10^{6}$ & 155 & 1,249 \\
  GeForce GTX 1080Ti & $51.11 \cdot 10^{6}$ & 175 & 999 \\
  RX Vega 64 & $44 \cdot 10^{6}$ & 230 & 399 \\
  GeForce RTX 2080 & $41 \cdot 10^{6}$ & 105 & 699 \\
  GeForce GTX TITAN X & $40 \cdot 10^{6}$ & 250 & 1,099 \\
  P104-100 & $38.89 \cdot 10^{6}$ & 127 & 569 \\
  RX Vega 56 & $38.75 \cdot 10^{6}$ & 210 & 339 \\
  GeForce RTX 2070 & $38.5 \cdot 10^{6}$ & 140 & 499 \\
  GeForce GTX 1080 & $34.07 \cdot 10^{6}$ & 121 & 633 \\
  RX 580 & $31.3 \cdot 10^{6}$ & 110 & 185 \\
  GeForce GTX 1070 & $31.1 \cdot 10^{6}$ & 108 & 319 \\
  GeForce GTX 1070Ti & $30.83 \cdot 10^{6}$ & 107 & 489 \\
  RX 570 & $29.85 \cdot 10^{6}$ & 65 & 142 \\
  RX 480 & $29.71 \cdot 10^{6}$ & 70 & 237 \\
  RX 470 & $29 \cdot 10^{6}$ & 60 & 340 \\
  GeForce GTX 1060 (6GB) & $23.81 \cdot 10^{6}$ & 95 & 264 \\
  GeForce GTX 1060 (3GB) & $19.32 \cdot 10^{6}$ & 69 & 189 \\
  GeForce GTX 1050Ti & $13.18 \cdot 10^{6}$ & 75 & 169 \\
  \hline
  \multicolumn{4}{|c|}{\textbf{Litecoin}} \\
  \hline
  Name & Hashes / s & Watt & Price (USD) \\
  \hhline{|=|=|=|=|}
  A6 LTCMaster & $123 \cdot 10^{7}$ & 1,500 & 3,000 \\
  A4+ LTCMaster & $62 \cdot 10^{7}$ & 750 & 1,500 \\
  Apollo LTC Pod & $10 \cdot 10^{7}$ & 100 & 299 \\
  \hline
  \hline
  \multicolumn{4}{|c|}{\textbf{Monero}} \\
  \hline
  Name & Hashes / s & Watt & Price (USD) \\
  \hhline{|=|=|=|=|}
  Shark Extreme 2 \small(8$ \times $AMD Vega 56) & 14,800 & 1,700 & 6,879 \\
  Shark Extreme 2 \small(6$ \times $AMD Vega 56) & 11,000 & 1,275 & 5,680 \\
  Shark Extreme 2 \small(8$ \times $AMD RX 580) & 6,880 & 1,100 & 4,590 \\
  Shark Extreme 2 \small(6$ \times $AMD RX 580) & 5,160 & 825 & 3,890 \\
  Shark Extreme 2 \small(4$ \times $AMD RX 580) & 3,440 & 550 & 2,590 \\
  RX Vega 64 & 2,020 & 140 & 399 \\
  RX Vega 56 & 1,920 & 140 & 339 \\
  GeForce RTX 2080Ti & 1,200 & 150 & 1,249 \\
  RX 580 & 976 & 89 & 185 \\
  RX 480 & 965 & 140 & 237 \\
  Ryzen Threadripper 1920X & 955 & 140 & 435 \\
  GeForce RTX 2080 & 898 & 132 & 699 \\
  GeForce GTX 2070 & 880 & 140 & 499 \\
  RX 470 & 840 & 120 & 340 \\
  GeForce GTX 1070 & 777 & 112 & 319 \\
  RX 570 & 740 & 90 & 142 \\
  Ryzen 7 2700X & 715 & 105 & 309 \\
  Ryzen 5 1600X & 532 & 47 & 179 \\
  Ryzen 5 1600 & 531 & 65 & 159 \\
  \hline
  \multicolumn{4}{|c|}{\textbf{Decred}} \\
  \hline
  Name & Hashes / s & Watt & Price (USD) \\
  \hhline{|=|=|=|=|}
  Whatsminer D1 & $44 \cdot 10^{12}$ & 2,200 & 1,588 \\
  Whatsminer DCR & $44 \cdot 10^{12}$ & 2,200 & 1,890 \\
  Antminer DR5 & $35 \cdot 10^{12}$ & 1,610 & 1,282 \\
  STU-U1+ & $12.8 \cdot 10^{12}$ & 1,850 & 1,560 \\
  STU-U1 & $11 \cdot 10^{12}$ & 1,600 & 1,389 \\
  \hline
\caption{Machines used in experiments}
\label{tbl:machines}
\end{longtable}

\end{document}